\documentclass[a4paper,fleqn,usenatbib]{mnras}

\usepackage{newtxtext,newtxmath}

\usepackage[T1]{fontenc}
\usepackage{ae,aecompl}

\usepackage{graphicx}	
\usepackage{amsmath}	
\usepackage{amssymb}	
\usepackage{ulem}

\newcommand{\Lsun}{\,$L_{\odot}$\,}	
\newcommand{\Msun}{\,$M_{\odot}$\,}	
\newcommand{\Prot}{\,$P_{\rm rot}$\,}	
\newcommand{\vrad}{\,$v_{\rm rad}$\,}	
\newcommand{\vsini}{\,$v \sin i$\,}	
\newcommand{\Teff}{\,$T_{\rm eff}$\,}	
\newcommand{\mps}{m s\,$^{-1}$\,} 
\newcommand{\logg}{\,$\log(g)$\,} 
\newcommand{\halpha}{\,H$\alpha$\,} 

\newcommand{\RCadd}[1]{\textcolor{blue}{\textbf{#1}}}

\title[TWA 9A and V1095 Sco. ]{The Surface Magnetic Activity of the Weak-Line T Tauri Stars TWA 9A and V1095 Sco}

\author[B. A. Nicholson et al.]{
B. A. Nicholson,$^{1,2}$\thanks{E-mail: belinda.nicholson@usq.edu.au}
G. A. J. Hussain,$^{2,3,4}$
J.-F. Donati,$^{3,4}$
C. P. Folsom,$^{3,4}$
M. Mengel,$^{1}$\and
B. D. Carter,$^{1}$
D. Wright,$^{1}$
and the MaTYSSE collaboration
\\
$^{1}$University of Southern Queensland, Computational Engineering and Science Research Centre, Toowoomba, Australia\\
$^{2}$European Southern Observatory, Karl Schwarzschild Str. 2, 85748 Garching, Germany\\
$^{3}$Univ. de Toulouse, UPS-OMP, IRAP, 14 av Belin, F–31400 Toulouse, France \\
$^{4}$CNRS, IRAP / UMR 5277, 14 av Belin, F–31400 Toulouse, France
}

\date{Accepted 2018 July 16. Received 2018 July 09; in original form 2018 May 04}

\pubyear{2018}

\begin{document}
\label{firstpage}
\pagerange{\pageref{firstpage}--\pageref{lastpage}}
\maketitle

\begin{abstract}
We present a detailed analysis of high-resolution spectropolarimetric observations of the weak-line T Tauri stars (wTTSs) TWA 9A and V1095 Sco as part of a wider survey of magnetic properties and activity in weak-line T Tauri stars, called MaTYSSE (Magnetic Topologies of Young Stars and the Survival of close-in giant Exoplanets). Our targets have similar masses but differing ages which span the stage of radiative core formation in solar-mass stars.  We use the intensity line profiles to reconstruct surface brightness maps for both stars. The reconstructed brightness maps for TWA 9A and V1095 Sco were used to model and subtract the activity-induced jitter, reducing the RMS in the radial velocity measurements of TWA 9A by a factor of $\sim 7$, and for V1095 Sco by a factor of $\sim 3$. We detect significant circular polarisation for both stars, but only acquired a high quality circular polarisation time-series for TWA 9A. Our reconstructed large-scale magnetic field distribution of TWA 9A indicates a strong, non-axisymmetric field. We also analyse the chromospheric activity of both stars by investigating their H$\alpha$ emission, finding excess blue-ward emission for most observations of V1095 Sco, and symmetric, double-peaked emission for TWA 9A, with enhanced emission at one epoch likely indicating a flaring event.



\end{abstract}



\section{Introduction}
Magnetic fields play an important role in the evolution of stars, particularly in the pre-main sequence stages. As a star evolves from a protostar obscured by a cloud of dust, it emerges as a young classical T Tauri Star (cTTS), accreting matter from a surrounding disc of gas and dust. Magnetic fields are understood to control this accreting matter, and dissipate excess angular momentum that would otherwise cause the star to break up as it forms \citep{Bouvier2013}. Once accretion has stopped and the inner disc has been cleared of gas and dust, the star is classed as a weak-line T Tauri star (wTTS), so called because of a lack of accretion-driven\halpha emission in its spectra. Given that the distinct populations of wTTSs and cTTSs span a similar range of ages and internal structure development, magnetic field studies can be used to investigate intrinsic physical differences between the accreting cTTSs and non-accreting wTTSs, and determine the effect, if any, that accretion has on the large-scale magnetic fields of these stars. 

The magnetic fields of a number of cTTSs have been studied in detail through the Magnetic Protostars and Planets (MaPP) project \citep[e.g.][]{Donati2007,Donati2008,Donati2010,Donati2011,Donati2012,Hussain2009}. These show a range in the large-scale surface magnetic field topologies, which appears to correlate with internal structure evolution: fully convective cTTSs show predominantly dipolar and axisymmetric fields, whereas cTTSs that have evolved to form a radiative core demonstrate complex multipolar, non-axisymmetric fields (see \cite{Gregory2012} for a more detailed discussion).  

In recent years, various wTTSs (V410 Tau \citep{Skelly2010}, LkCa 4 \citep{Donati2014}, V819 Tau\citep{Donati2015}, V830 Tau \citep{Donati2015,Donati2016,Donati2017}, TAP 26 \citep{Yu2017}, Par 1379 and Par 2244 \citep{Hill2017}) have had their large-scale surface magnetic fields mapped thanks to the Magnetic Topologies of Young Stars and the Survival of close-in giant Exoplanets (MaTYSSE) large programme, an international, multi-telescope programme (CFHT, TBL and ESO 3.6m) aimed at gathering spectropolarimetric observations of wTTSs. These studies have found an even larger range of magnetic topologies than was found in accreting stars; from predominantly dipolar, poloidal field in the case of V830 Tau, to a predominantly toroidal field configuration in the case of LkCa 4, implying that changing the accretion state may have an impact on surface field topology. 

In addition to characterising their surface features, the MaTYSSE programme also aims to search for close-in giant exoplanets around wTTSs, applying the surface brightness information from Doppler Imaging (DI) to aid in reducing activity-induced jitter in the radial velocity curves.  WTTSs are ideal targets for searching for young exoplanets, due to the lack of interference from accretion or obscuration by a disc, and their fast rotation rates making them suitable DI targets. With this jitter-reduction technique there have been two close-in, giant, exoplanets detected, one each around V830 Tau and TAP 26. 

In this paper we present an analysis of spectropolarimetric data for two wTTSs, TWA 9A and V1095 Sco. We reconstruct the surface brightness topologies of both TWA 9A and V1095 Sco, and for TWA 9A we reconstruct the surface large-scale magnetic field. We also present an analysis of radial velocity curves for both stars using one of the activity filtering techniques developed in previous MaTYSSE studies, and compare the effectiveness of different DI techniques on activity jitter reduction. TWA 9A and V1095 Sco are presented together as both of them were observed with the HARPS spectrograph, and provide an interesting contrast in rotation rates and evolutionary stages, while having similar spectral types, and can both be considered young solar analogues. TWA 9A is a well-characterised wTTS in the TW Hya association, with numerous existing observations from optical to X-ray wavelengths. V1095 Sco has also been the target of multi-wavelength studies, but has not been characterised in as much detail. 

In Section \ref{sec:obs} we detail the observations, and in Section \ref{sec:starpar} we determine the parameters and evolutionary statuses of our stars from the literature and our own calculations. Section \ref{sec:TomMod} describes the process of creating the surface brightness and magnetic field maps, and we examine the\halpha emission in the spectra of both stars in Section \ref{sec:halpha}. Section \ref{sec:RVs} presents analysis of the radial velocity curves of each star, and Section \ref{sec:periods} explores the stellar rotation period measurements in our work. We summarise and discuss our findings in Section \ref{sec:discussion}.

\section{Observations}
\label{sec:obs}
TWA 9A and V1095 Sco were both observed as part of the MaTYSSE programme using the HARPS spectropolarimeter on the 3.6m ESO Telescope at the La Silla Observatory in Chile. The journals of observations for TWA 9A and V1095 Sco are shown in Tables \ref{tab:obs_twa9a} and \ref{tab:obs_v1095sco}, respectively. A total of 14 observations were made of TWA 9A over 17 nights, and 13 observations of V1095 Sco over 13 nights, corresponding to approximately three rotations of TWA 9A, and over four rotations of V1095 Sco. Each observation was taken as a set of 4 sub-exposures in alternating configurations of the quarter-wave retarder to remove to first-order any spurious polarisation signals. The wavelength coverage of the spectra is from 383 to 691 nm with a spectral resolution of $\sim115000$, and peak signal-to-noise ratios (SNR) ranging between 36 and 116 for TWA 9A, and 60 and 100 for V1095 Sco.

The data were reduced using {\sc Libre Esprit}, a pipeline reduction tool designed for spectropolarimetric observations, with adaptations for use with HARPS polarimetric data \citep{Hebrard2016}, following the procedure outlined by \cite{Donati1997}. 

\begin{table*}
	\centering
	\caption{Journal of observations for star TWA 9A. This table lists 1) the date of observation, 2) the Heliocentric Julian date, 3) the exposure time as a set of sub-exposures, 4) the stellar rotational phase based on a period of 5.01 days, with the zero-point set as the date of the first observation, 5) the peak SNR in Stokes I per observation, 6) the peak SNR in Stokes V per observation, 7) the SNR after performing least squares deconvolution (LSD, see Section \ref{sec:LSD}) in Stokes I, and 8) the SNR after LSD in Stokes V.  }
	
	\label{tab:obs_twa9a}
	\begin{tabular}{cccccccccc} 
		\hline
		Date   & HJD        & Exposure  & Rotation  & Stokes I & Stokes V & Stokes I & Stokes V \\
		(2014) & (2456000+) &  Time (s) & Phase     & Obs. SNR & Obs. SNR & LSD SNR  & LSD SNR \\
		\hline
		 23 May   & 801.64194 & $4\times1100$ & 0.00 & 37 & 32 & 798  & 1731 \\
		 25 May   & 803.59193 & $4\times1100$ & 0.39 & 56 & 58 & 1036 & 2873 \\
		 26 May   & 804.58780 & $4\times1100$ & 0.59 & 60 & 56 & 1057 & 3190 \\
		 27 May   & 805.58943 & $4\times1100$ & 0.79 & 41 & 37 & 1008 & 2894 \\
		 30 May   & 808.59376 & $4\times1700$ & 1.39 & 82 & 73 & 1090 & 4225 \\
 		 31 May   & 809.51753 & $4\times1700$ & 1.57 & 93 & 83 & 1109 & 5161 \\
		 01 June  & 810.53135 & $4\times1700$ & 1.77 & 92 & 84 & 1126 & 5178 \\		 
		 02 June  & 811.52101 & $4\times1700$ & 1.97 &116 &103 & 1151 & 6670 \\
		 03 June  & 812.58791 & $4\times1700$ & 2.18 & 49 & 45 & 856  & 2256 \\
		 05 June  & 814.55552 & $4\times1700$ & 2.58 & 76 & 70 & 1067 & 4156 \\
		 06 June  & 815.66190 & $4\times1700$ & 2.80 & 52 & 46 & 845  & 2454 \\
		 07 June  & 816.55456 & $4\times1700$ & 2.98 & 76 & 66 & 1083 & 4075 \\
		 08 June  & 817.54804 & $4\times1700$ & 3.17 & 76 & 71 & 1090 & 4239 \\
		 09 June  & 818.54821 & $4\times1700$ & 3.37 & 83 & 75 & 1120 & 4732 \\
		\hline
	\end{tabular}
\end{table*}

\begin{table*}
	\centering
	\caption{Table of observations for star V1095 Sco. For description, see caption of Table \ref{tab:obs_twa9a}. A rotation period of 2.90 days was used for calculating the rotation phase.}
	\label{tab:obs_v1095sco}
	\begin{tabular}{cccccccccc} 
		\hline
		Date   & HJD        & Exposure  & Rotation  & Stokes I & Stokes V & Stokes I & Stokes V \\
		(2014) & (2456000+) &  Time (s) & Phase     & Obs. SNR & Obs. SNR & LSD SNR  & LSD SNR \\
		\hline
		 14 June & 823.63745 & $4\times1250$ & 0.00 & 67 & 62 & 839 & 2310 \\
		 15 June & 824.65217 & $4\times1250$ & 0.35 & 61 & 56 & 846 & 2168 \\
		 16 June & 825.72489 & $4\times1250$ & 0.72 & 67 & 61 & 877 & 2393 \\
		 17 June & 826.68254 & $4\times1900$ & 1.05 & 85 & 76 & 914 & 3190 \\
		 18 June & 827.68058 & $4\times1900$ & 1.39 & 85 & 80 & 905 & 3166 \\
 		 19 June & 828.71628 & $4\times1900$ & 1.75 & 87 & 78 & 935 & 3358 \\
		 20 June & 829.62037 & $4\times1900$ & 2.06 & 81 & 75 & 913 & 2910 \\		 
		 21 June & 830.59708 & $4\times1900$ & 2.40 & 64 & 38 & 846 & 1363 \\
		 22 June & 831.64483 & $4\times1900$ & 2.76 & 92 & 78 & 980 & 3129 \\
		 23 June & 832.59325 & $4\times1900$ & 3.09 & 98 & 91 & 967 & 3605 \\
		 24 June & 833.59082 & $4\times1900$ & 3.43 & 68 & 64 & 828 & 2397 \\
		 25 June & 834.62734 & $4\times1900$ & 3.79 & 63 & 59 & 838 & 2233 \\
		 26 June & 835.59555 & $4\times1900$ & 4.12 & 67 & 63 & 771 & 2386 \\
		\hline
	\end{tabular}
\end{table*}

\subsection{Least Squares Deconvolution}
\label{sec:LSD}
To increase the SNR in our intensity and Stokes V line profiles, absorption lines are combined using Least-Squares Deconvolution \citep[LSD,][]{Donati1997}. This process involves de-convolving the observed stellar spectrum with an absorption line list, giving the profile of an average line (called an LSD profile) with an enhanced SNR compared to a single line profile in the original spectrum. In order to create this average profile, we first create a line list specifying the depths and Land\'e factors of selected photospheric lines based on the star's spectral type. For both stars, a list was generated using the VALD3 database \citep{Ryabchikova2015}, using the stars' respective effective temperatures (\Teff) and log surface gravities ($\log g$) as inputs (determination of these quantities is detailed in Section \ref{sec:spec_type}). These lists were edited to remove lines that were contaminated by obvious emission features, strong stellar absorption features, and telluric absorption features, or had a line depth relative to the deepest line of less than 0.1. For each star, the LSD line profiles were renormalised and scaled by the mean equivalent width of the set of LSD profiles. The resulting LSD profiles have a mean SNR in Stokes V (Stokes I) of 3845 (1031) for TWA 9A and 2662 (881) for V1095 Sco. The peak SNR in Stokes V for all observations for both stars is given in Tables \ref{tab:obs_twa9a} and \ref{tab:obs_v1095sco}.

\section{Evolutionary states of TWA 9A and V1095 Sco}
\label{sec:starpar}
\subsection{Spectral classification}
\label{sec:spec_type}
Determining the spectral type of a PMS star is not straightforward, and appears be sensitive to the procedure used. This is exemplified by the range in spectral types and temperatures found in the literature and determined here for both stars. For TWA 9A, the literature values for spectral type range between K5 \citep[e.g.][]{Manara2013} and K7 \citep[e.g.][]{Pecaut2013}, cooler than the K4 we determine here.  Manara et al. determine the spectral types from the relative depths of several molecular bands using low-resolution, visible spectra in the range 580 nm to 900 nm. Pecaut \& Mamajek use low resolution spectra in a narrower range of wavelengths (560 to 690 nm), and compare several atomic lines, blends and molecular bands to a set of PMS spectral standards. For cooler stars, the inclusion of cooler molecular bands and blends for spectral typing leads to a cooler temperature, as these features will be present in the spots on the surface, skewing the photospheric temperature. The exception to this is the use of TiO molecular bands, which \cite{Herczeg2014} find to be poorly modelled for PMS stars, giving a hotter spectral type compared to when they are excluded. This potentially explains the difference in spectral types between Manara et al. who include TiO and Pecaut \& Mamajek who do not. 

Given the uncertainty in the literature, we re-determine the \Teff and $\log g$ of both stars using the spectral typing procedure outlined in \cite{Donati2012}, which follows the method of \cite{Valenti2005} of comparing high resolution optical spectra to a grid of synthetic spectra generated for a range of \Teff and \logg values. For TWA 9A we find a \Teff of $4400\pm50$ K and $\log g$ of $4.1\pm0.2$. Using the conversion table of \cite{Herczeg2014}, this corresponds to a spectral type around K4.  This spectral typing procedure uses only atomic lines that are less sensitive down to spot temperatures. 

For V1095 Sco we find a \Teff of $4350\pm50$ K and $\log g$ of $3.8\pm0.2$, which also corresponds to a spectral type of K4. \cite{Krautter1997} found a spectral type of K5 using low-resolution spectra, and comparing with MS spectral standards from the MK star catalogue \citep{Buscombe1981,Buscombe1988}. \cite{Gregory2016} find that, at a given temperature, the use of MS spectral standards gives a systematically cooler spectral type for mid-G to mid-K stars. This may explain the slightly cooler spectral type found by Krautter et al. compared with our determined spectral type.

\subsection{Determination of stellar parameters and inclination}
\label{sec:Lum}
In order to determine the most accurate surface maps, we must first determine the inclination of our stars with respect to the observer. This is evaluated based on rotation period, \vsini  and radius of the star. The radius, along with mass and age, is determined from PMS stellar evolution tracks, in this case the models of \cite{Baraffe2015}, using \Teff (See Section \ref{sec:spec_type}) and luminosity. 

There are a variety of values of luminosity for TWA 9A in the literature, ranging from $0.13L_{\odot}$ \citep{Mcdonald2012} up to $0.34 L_{\odot}$ \citep{Pecaut2013}. The variation in these values can be at least partly accounted for by differences in the distance used. McDonald et al. use the Hipparcos-derived parallax distance of 46 pc for their luminosity calculation. Pecaut \& Mamajek redetermine their own distance ($70.0 \pm 3.8$ pc) using the kinematic parameters from  \cite{Weinberger2013}, and the method described in \cite{Mamajek2005}. \cite{Manara2013} use the distance of $\sim68$ pc from \cite{Mamajek2005}, but assume a different apparent magnitude, and so find a lower luminosity of $0.25$ $L_{\odot}$. All luminosity calculations mentioned in the literature take the interstellar extinction, $A_v$, to be zero. This likely results from using low resolution spectra to determine extinction, which can lead to zero, or negative extinction values, as in the case of \cite{Herczeg2014}. We re-determine the luminosity using the parallax measurement from {\it Gaia} Data Release 2 \citep[DR2,][]{GaiaCollab2016,GaiaDR22018}, which gives a distance of $76.4 \pm 0.3$ pc. We calculate extinction, $A_v$ using apparent V-band magnitude, $m_V=11.143\pm0.029$ mag, from \cite{Henden2016}, V-band bolometric correction $BC_V=-0.77\pm0.18$ and intrinsic colour $(B-V)_0=1.1\pm0.09$ from from \cite{Pecaut2013}, and observed colour $(B-V)=1.244 \pm 0.067$ from Henden et al., giving $A_v = 0.42 \pm 0.35$ mag. This gives a luminosity of $0.48 \pm 0.09$ $L_{\odot}$. This luminosity is higher than any listed in the literature. This higher luminosity is expected given our larger distance, and high value of $A_v$, which could be due to reddening from the presence of photospheric spots, rather than interstellar dust. 

The literature values for the luminosity of V1095 Sco vary widely from $0.74$ $L_{\odot}$ \citep{Sartori2003} to $3.45$ $L_{\odot}$ \citep{Wahhaj2010}. These differences can also be attributed to differences in distance and extinction corrections used. Since the distance to V1095 Sco had not been measured at the time, both authors use the distance to the Lupus star forming region. Sartori et al. takes this distance to be 147 pc from \cite{Bertout1999}, whereas Wahhaj et al. use 200 pc from \cite{Comeron2008}. Both authors differ in their choice of extinction, with Sartori et al. using $A_v=0.0$, and Wahhaj et al. finding $A_v=0.7$. These stark differences in extinction and distance values, as well as subtle differences in values of apparent magnitude, account for the large differences in luminosity. We recalculate the luminosity for V1095 Sco using the parallax measurement from Gaia DR2, which give a distance of $162 \pm 1$ pc. We take the observed colour $(B-V)=1.309 \pm 0.045$ and apparent magnitude $V=11.424\pm0.041$ from \cite{Henden2016}, and intrinsic colour $(B-V)_0 = 1.11\pm0.09$ and bolometric correction $BC = -0.77\pm0.18$ from \cite{Pecaut2013} and calculate an extinction $A_v=0.62\pm 0.31$, giving a luminosity of $2.0 \pm 0.3$ $L_\odot$. A higher estimate of $A_v$ compared to TWA 9A is expected given the greater distance to this star, and is likely affected by both interstellar dust and reddening due to spots. 

\begin{figure*}
\centering
\includegraphics[width=1.0\textwidth]{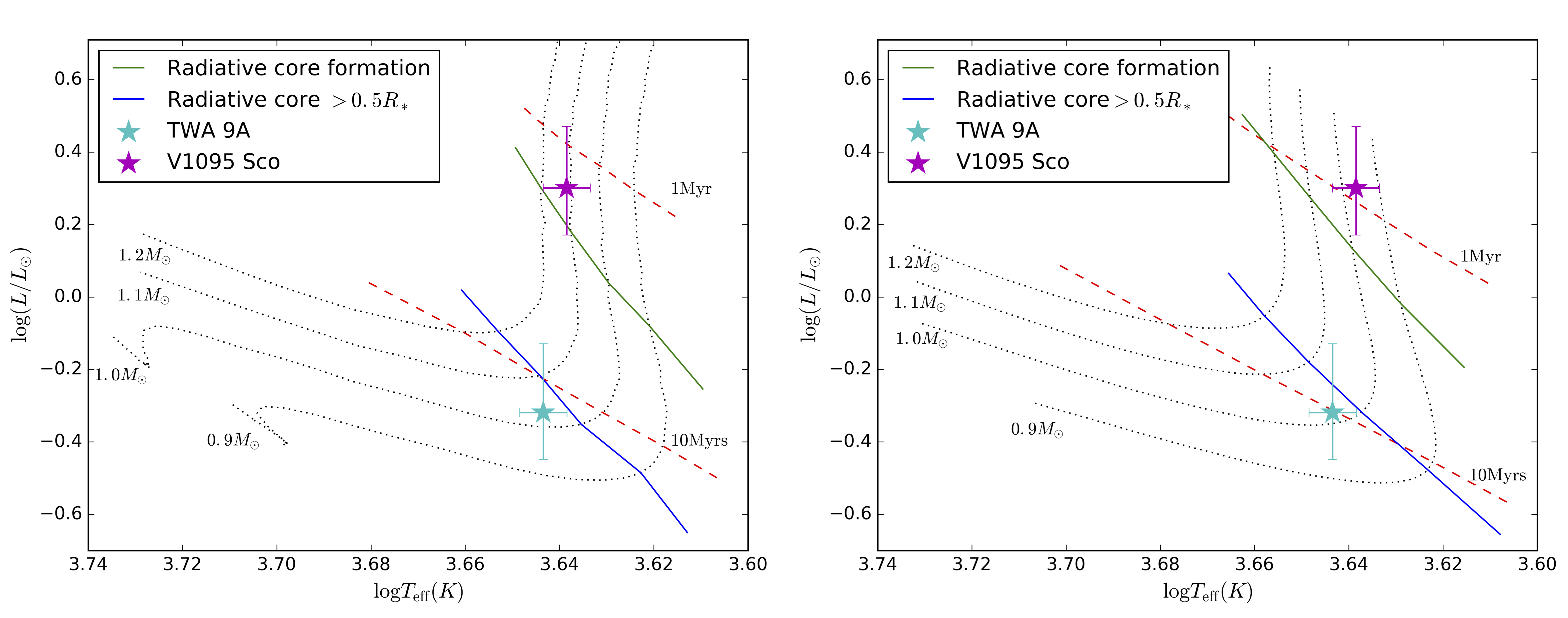}
\caption{HR diagrams showing the locations, including $1\sigma$ uncertainties, of TWA 9A (cyan) and V1095 Sco (magenta) with relation to the \citet{Baraffe2015} (right), and \citet{Siess2000} (left) PMS evolutionary models. In both figures the black dotted lines are evolutionary tracks for 0.9\Msun, 1.0\Msun, 1.1\Msun and 1.2\Msun, and the red dashed lines are isochrones for 1 Myr and 10 Myrs. The green line indicates the boundary between full convection and the formation of a radiative core, and the blue line indicated the point at where the radiative core becomes larger than $0.5 R_*$.}
\label{fig:HRdiagramBoth}
\end{figure*}

Figure \ref{fig:HRdiagramBoth} shows the locations of TWA 9A and V1095 Sco on the HR diagram with respect to the PMS evolutionary models of \cite{Baraffe2015} (left), and, for comparison with other MaTYSSE stars, the models of \cite{Siess2000} (right). Our mass, radius and age estimates for our stars are based on the models of \cite{Baraffe2015}, and are shown in Table \ref{tab:stellar_param}. These models largely agree for stars on the Henyey track, as is the case of TWA 9A, but there are discrepancies for those on the Hayashi track, with a difference in mass estimate of $\sim 0.1$ \Msun for V1095 Sco. However, this difference is within the $3\sigma$ uncertainty in temperature for V1095 Sco. Examining the proximity of these stars on the HR diagram with respect to the point of radiative core formation (green line), and the formation of a substantial radiative core ($> 0.5R_*$, blue line), both models indicate that V1095 Sco is still fully convective, but close to forming a radiative core. TWA 9A, however, is estimated to have already formed a substantial radiative core ($0.60\pm0.06R_*$), and is more evolved than V1095 Sco. 
 
We calculate the inclination of each star with respect to the observer\footnote{In this convention, $i = 0^{\circ}$ means a pole-on view, and $i = 90^{\circ}$ means an equator-on view} from the radii, \vsini's and rotation periods given in Table \ref{tab:stellar_param}. For V1095 Sco, we estimate an inclination of $\sim 45^{\circ}$, slightly higher than the inclination of $\sim 40^{\circ}$ that results from the luminosity value of $3.45$ $L_{\odot}$ from Wahhaj et al.  For TWA 9A we find an inclination of $\sim 65^{\circ}$. It is interesting to note that none of the literature luminosity values for TWA 9A result in a radius that gives a viable $\sin(i)$ value (i.e. $\sin(i)<1$), supporting our larger luminosity determination.

\begin{table}
	\centering
	\caption{Summary of the properties of TWA 9A and V1095 Sco. Uncertainties are given where available.  }
	\label{tab:stellar_param}
	\begin{tabular}{lcc}
		\hline
		 & TWA 9A & V1095 Sco.\\
		\hline
		Mass, $M_*$ (\Msun) & $1.00\pm0.1$ & $0.96^{+0.05}_{-0.08}$ \\
		$R \sin i$, ($R_{\odot}$)& $1.070\pm0.003$ & $1.75\pm0.03$ \\ 
		Radius, $R_*$ ($R_{\odot}$) & $1.2^{+0.3}_{-0.2}$ & $2.5^{+0.4}_{-0.5} $ \\  
		Age (Myrs) & $10^{+13}_{-5}$ & $0.9^{+1.1}_{-0.4}$ \\
		Luminosity, $L_*$ (\Lsun) & $0.5 \pm 0.2$ & $2.0 \pm 0.7$ \\
		Distance, (pc) & $76.4 \pm 0.3$ & $162 \pm 1$ \\
		Extinction, $A_v$ (mag) & $0.42 \pm 0.35$ & $0.62 \pm 0.32$ \\
		\vsini (k\mps) & $10.8\pm0.02$ & $30.6\pm0.1$ \\
		Rotation Period, \Prot (d) & $5.01\pm0.01$ & $2.9\pm0.05$ \\ 
		Effective Temperature, \Teff (K) & $4400 \pm 50$ & $4350 \pm 50$ \\
		Surface Gravity, $\log(g)$ & $4.1\pm0.2$ & $3.8\pm0.2$\\
		Spectral Type  & K4 & K4\\
		Inclination Angle, $i$ (Degrees) & $\sim 65$ & $\sim 45$ \\
		\hline
	\end{tabular}
\end{table}

\section{Tomographic Modelling}
\label{sec:TomMod}
\subsection{Brightness Mapping}
\label{sec:DI}
Using the stellar parameters described in Section \ref{sec:starpar} and our Stokes I LSD profiles (Section \ref{sec:obs}), we reconstruct surface brightness maps of TWA 9A and V1095 Sco using the technique of Doppler Imaging (DI). For this we use an updated version of the code {\sc DoTS} \citep{CollierCameron1997}, which now reconstructs surface brightness, as described by \cite{Donati2014}, allowing for both dark and bright regions, rather than spot filling alone. We estimate the local line profile assuming a Milne-Eddington model atmosphere, invert the profiles and use a maximum entropy regularisation to reconstruct an image of surface brightness that requires the least amount of information to fit the observed data to a given level described by the reduced chi-squared, ${\rm \chi^2_r}$. 

\begin{figure}
\centering
\includegraphics[width=0.4\textwidth]{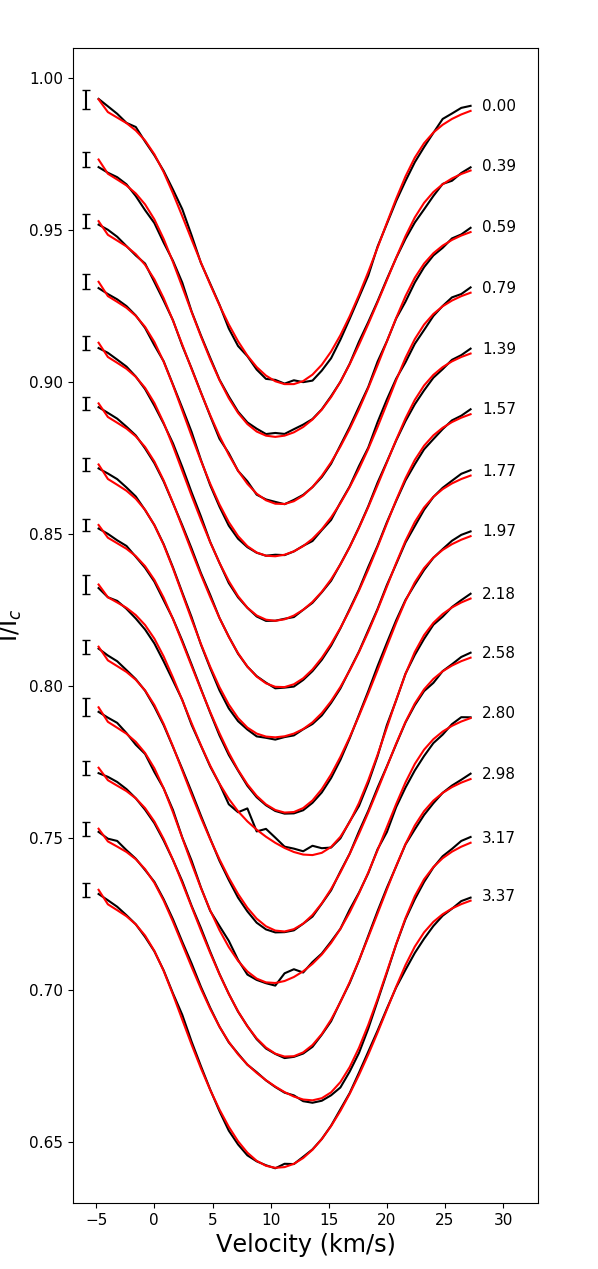}
\caption{These plot shows the Stokes I LSD profiles for TWA 9A (black), and the best fit synthetic line profile (red). The phases of observation are given to the right of each line, with the zero phase set as the first observation and our derived stellar rotation period $P_{\rm rot}=5.01$ days. The error bar on the left is the mean $3\sigma$ error for each Stokes I profile.}
\label{fig:StokesI_TWA}
\end{figure}

\begin{figure}
\centering
\includegraphics[width=0.4\textwidth]{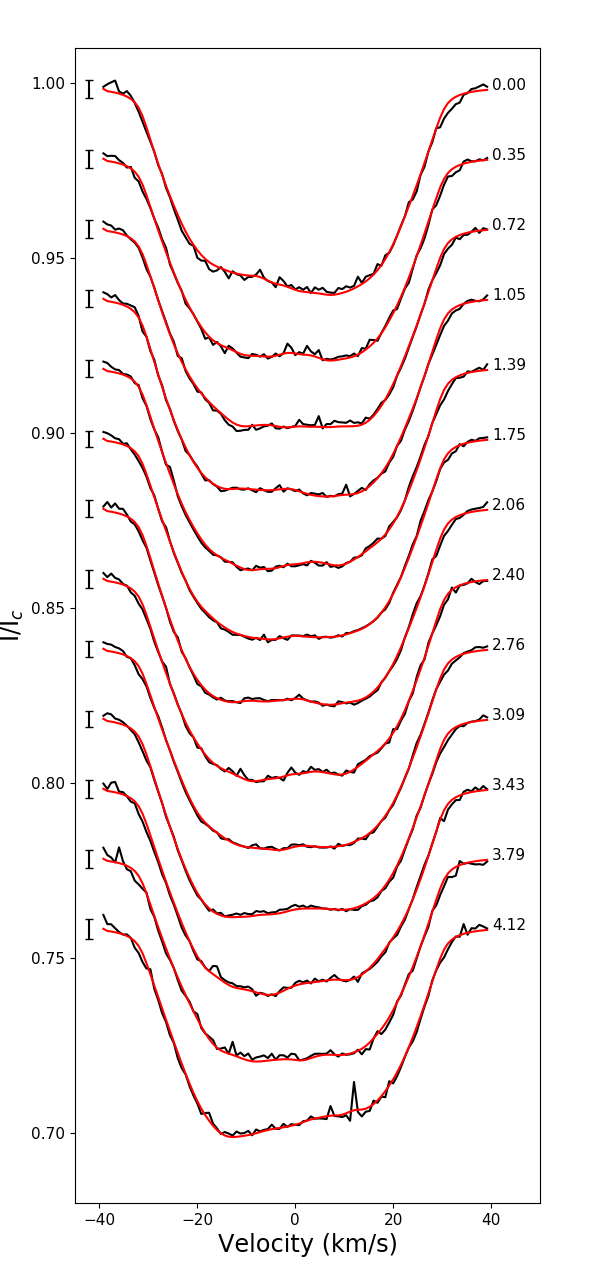}
\caption{These plot shows the Stokes I LSD profiles for V1095 Sco (black), and the best fit synthetic line profile (red). The phases of observation are given to the right of each line, with the zero phase set as the first observation and our derived stellar rotation period $P_{\rm rot}=2.90$ days. The error bar on the left is the mean $3\sigma$ error for each Stokes I profile.}
\label{fig:StokesI_RXJ}
\end{figure}

In the process of reconstructing the brightness map we optimise for the mean radial velocity \vrad, local line equivalent width (EW), \vsini, and rotation period, \Prot, for each star; the latter two of these parameters are given in Table \ref{tab:stellar_param}. The fits to the intensity profiles are shown by the red lines in Figures \ref{fig:StokesI_TWA} and \ref{fig:StokesI_RXJ}. 

\begin{figure}
\centering
\includegraphics[width=0.5\textwidth]{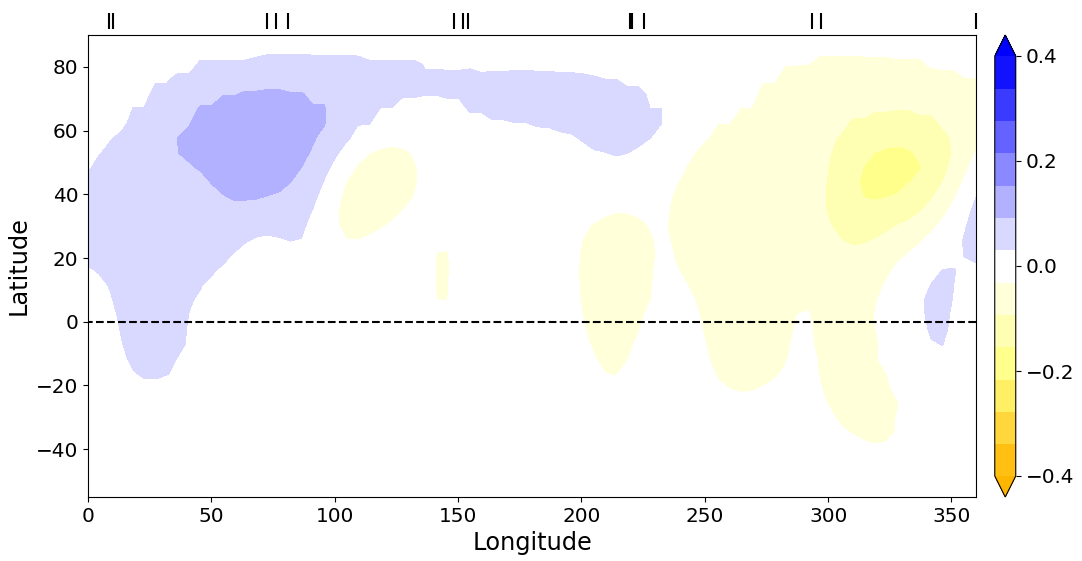}
\caption{Surface brightness map of TWA 9A, shown as a rectangular projection of longitude and latitude. The colour scale indicates log surface brightness relative to a photosphere (0.0), with blue indicating the presence of plage, and yellow the presence of spots. The observations are marked by black ticks across the top, and black dashed line indicates the equator.}
\label{fig:DITWA}
\end{figure}

\begin{figure}
\centering
\includegraphics[width=0.5\textwidth]{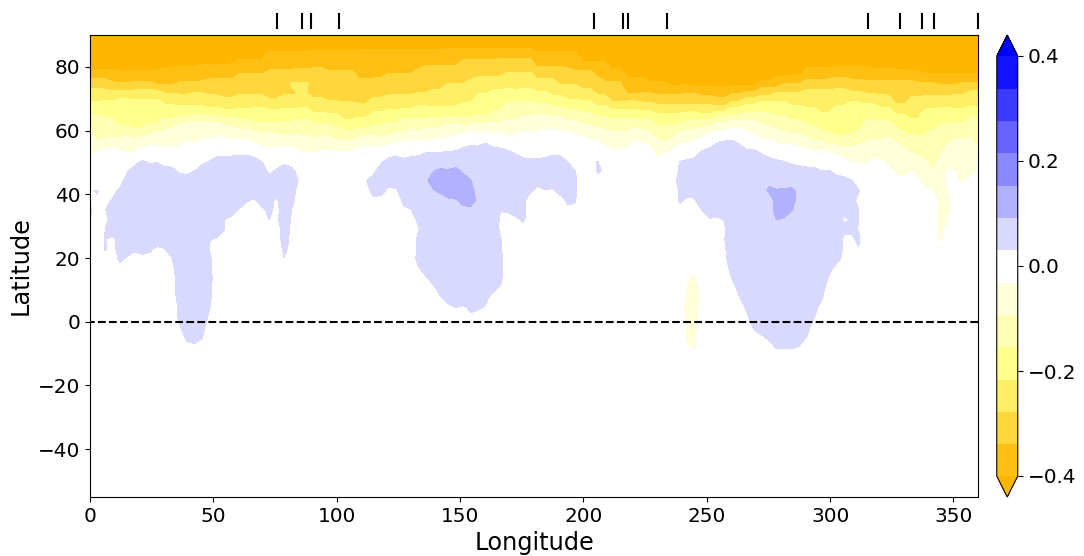}
\caption{Surface brightness map of V1095 Sco, shown as a rectangular projection of longitude and latitude. Plot description same as for Figure \ref{fig:DITWA}.}
\label{fig:DIV1095}
\end{figure}

The surface brightness maps of TWA 9A and V1095 Sco are shown in Figures \ref{fig:DITWA} and \ref{fig:DIV1095}, respectively. These plots are rectangular projections of the stellar surface in longitude and latitude. 

Our reconstructions show both dark and bright spotted regions. TWA 9A shows a complex surface brightness map, with strong signatures of both spots and plage. The spots are located near the equator and mid latitudes, whereas the bright plage is found at a higher latitude. 

The surface brightness map of V1095 Sco is simpler, with one large polar spot, and three areas of plage at mid-latitudes. Inspection of the intensity profiles of V1095 Sco show characteristic flat-bottomed profiles indicating the need for a dark polar spot in the resulting maps. The fact that the plage is preferentially associated with gaps in the phase coverage indicate that it could be spurious, though a better fit is achieved with the inclusion of plage than with spots alone.

\subsection{Magnetic Mapping}

\label{sec:ZDI}
\begin{figure*}
\centering
\includegraphics[width=1.0\textwidth]{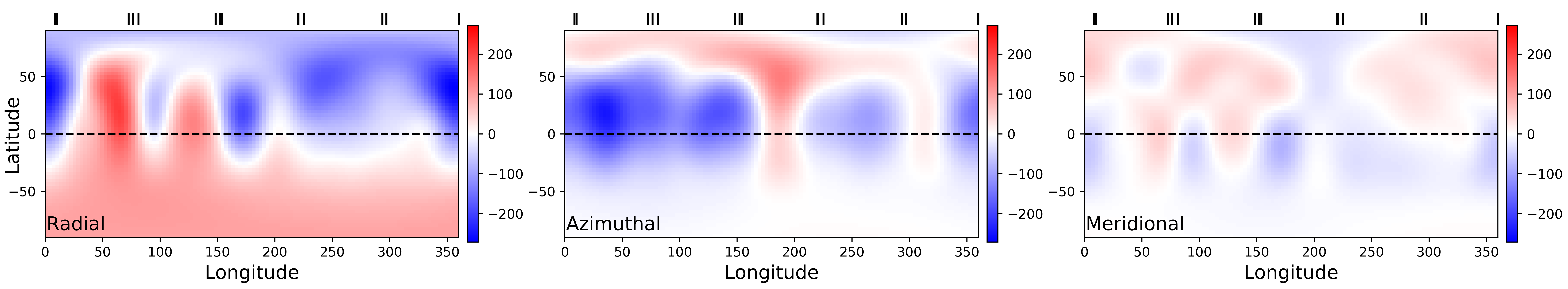}
\caption{Maps of the radial, azimuthal and Meridional magnetic field components for TWA 9A. Colour scale is magnetic field strength in Gauss, the dashed line indicates the equator, and tick marks across the top indicate observed longitudes.}
\label{fig:TWA9Amagmap}
\end{figure*}

By inverting the Stokes V profiles we can also reconstruct the large scale magnetic fields of TWA 9A. Our reconstructions are shown in Figure \ref{fig:TWA9Amagmap}. For this analysis we used the {\sc ZDIPy} code of \cite{Folsom2018}. This is based on the Zeeman Doppler Imaging (ZDI) method of \cite{Semel1989}, \cite{Donati1997} and \cite{Donati2006}, representing the field components as spherical harmonics and using the maximum entropy regularisation of \cite{Skilling1984} to solve for the best fit configuration. Its performance has been calibrated against the published code of \cite{Donati2006}. We used the same stellar parameters determined from our analysis of the intensity profiles (see Table \ref{tab:stellar_param}), and take into account the surface brightness variations in our reconstruction of the magnetic field. The Stokes I local line profile parameters are comparable to that used in the {\sc DoTS} code. The Stokes V profile were modelled using the weak field approximation with mean Land\'e factor and central wavelength corresponding to those used to extract the LSD profiles and for the brightness reconstruction. The optimum value of $\chi^2_r$ aim was chosen using the method of \cite{AlvaradoGomez2015}, who choose the optimum target $\chi^2_r$ using the maxima in the second derivative of entropy as a function of  $\chi^2_r$ in a set of converged magnetic field solutions. According to our analysis this implies the map obtained fitting the data at a $\chi^2_r$ of 2.0. This solution gives a mean magnetic field of 113 G, with a maximum of 296 G. Analysing the amount of energy in different spherical harmonic components of this magnetic field solution, we find a non-axisymmetric ($57\%$ energy in non-axisymmetric components), predominantly poloidal large scale field ($\sim 69\%$ poloidal, $\sim 31\%$ toroidal). In the poloidal component, the field is predominantly dipolar and quadrupolar, with only $\sim 11\%$ in octopolar or higher orders. The dipolar component of the field has a maximum intensity of $127$ G, and is tilted at an angle of $36.5$ degrees to the rotation axis, with the pole located at phase 0.26. For comparison, we calculate the magnetic field configuration using zDoTS \citep{Hussain2000,Hussain2016} with the same local line profiles used for our brightness mapping. The best-fit model also yields a predominantly non-axisymmetric field ($55\%$ in non-axisymmetric components), and a predominantly poloidal field ($\sim 64\%$ poloidal, $\sim 37\%$ toroidal). The poloidal field in this estimate is slightly more concentrated in the dipolar and quadrupolar components, with $8.5\%$ of the poloidal field energy being containing in octopolar components or higher. 

\begin{figure}
\centering
\includegraphics[width=0.3\textwidth]{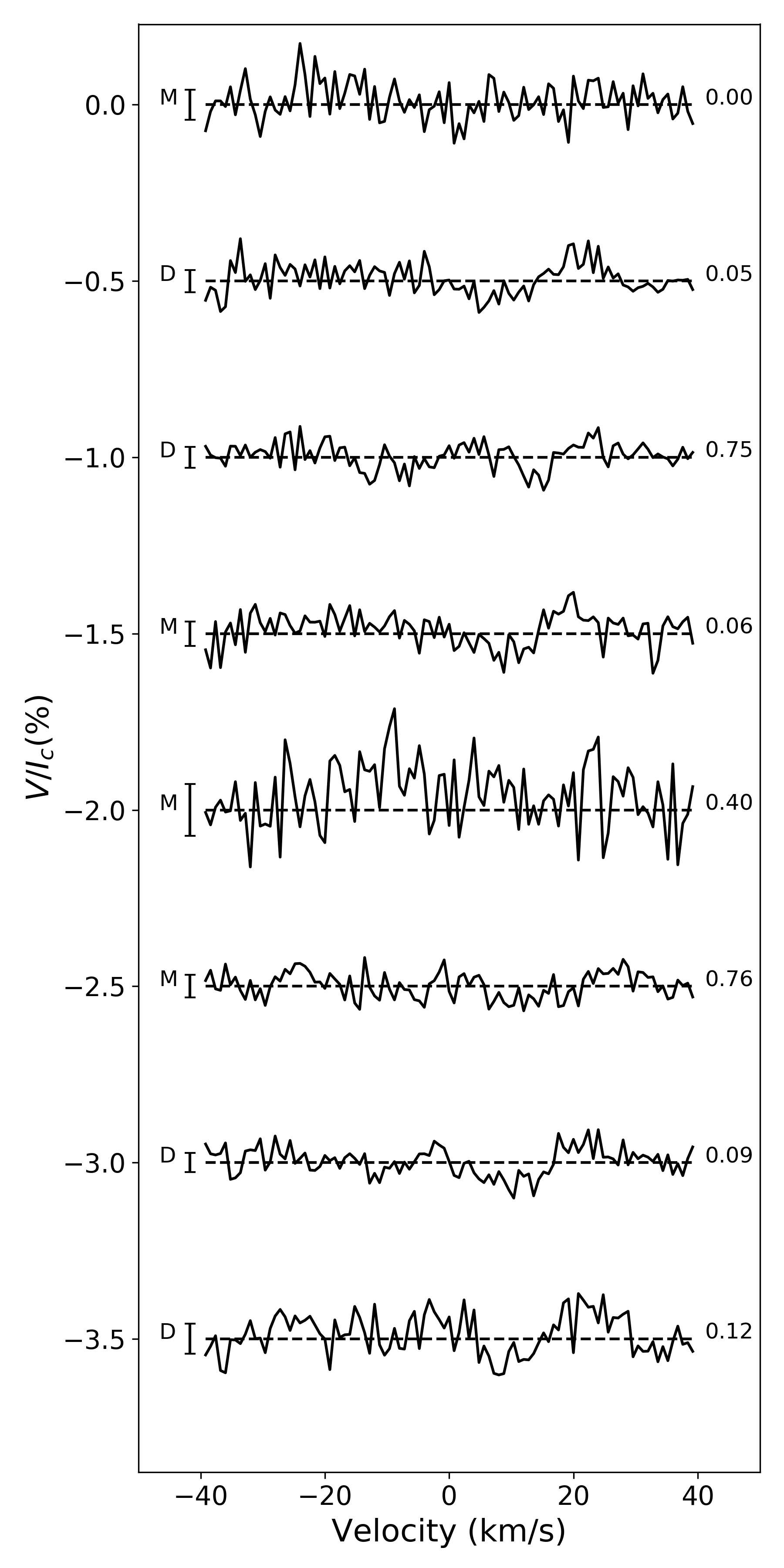}
\caption{Stokes V intensity data for V1095 Sco, expressed as a percentage of Stokes I continuum. The letter on the right of reach spectrum indicated the level of magnetic field detection: `M' for marginal detection (false alarm probability between $10^{-5}$ and $10^{-3}$, and `D' for a definite detection (false alarm probability less than $10^{-5}$). Spectra for which there is no detection in stokes V are not shown. Error bars on the left are the mean $1\sigma$ errors for each observation, and the numbers on the right indicate the phase of observation, based on a rotation period of 2.9 days. }
\label{fig:StokesV_RXJ}
\end{figure}

Magnetic mapping using the Stokes V LSD line profiles of V1095 Sco was attempted, but a reliable solution could not be found due to the poor SNR of the observations. The Stokes V spectra of observation for which there are marginal (false alarm probability between $10^{-5}$ and $10^{-3}$) or definite (false alarm probability less than $10^{-5}$) signal detections are plotted in Figure \ref{fig:StokesV_RXJ}. Comparing the Stokes V LSD profiles to the null for these observations suggests that the signal in some of these marginal detections is spurious.  The phase coverage in the remaining robustly detected signatures was insufficient to yield a reliable surface magnetic field map.


\begin{figure}
\centering
\includegraphics[width=0.3\textwidth]{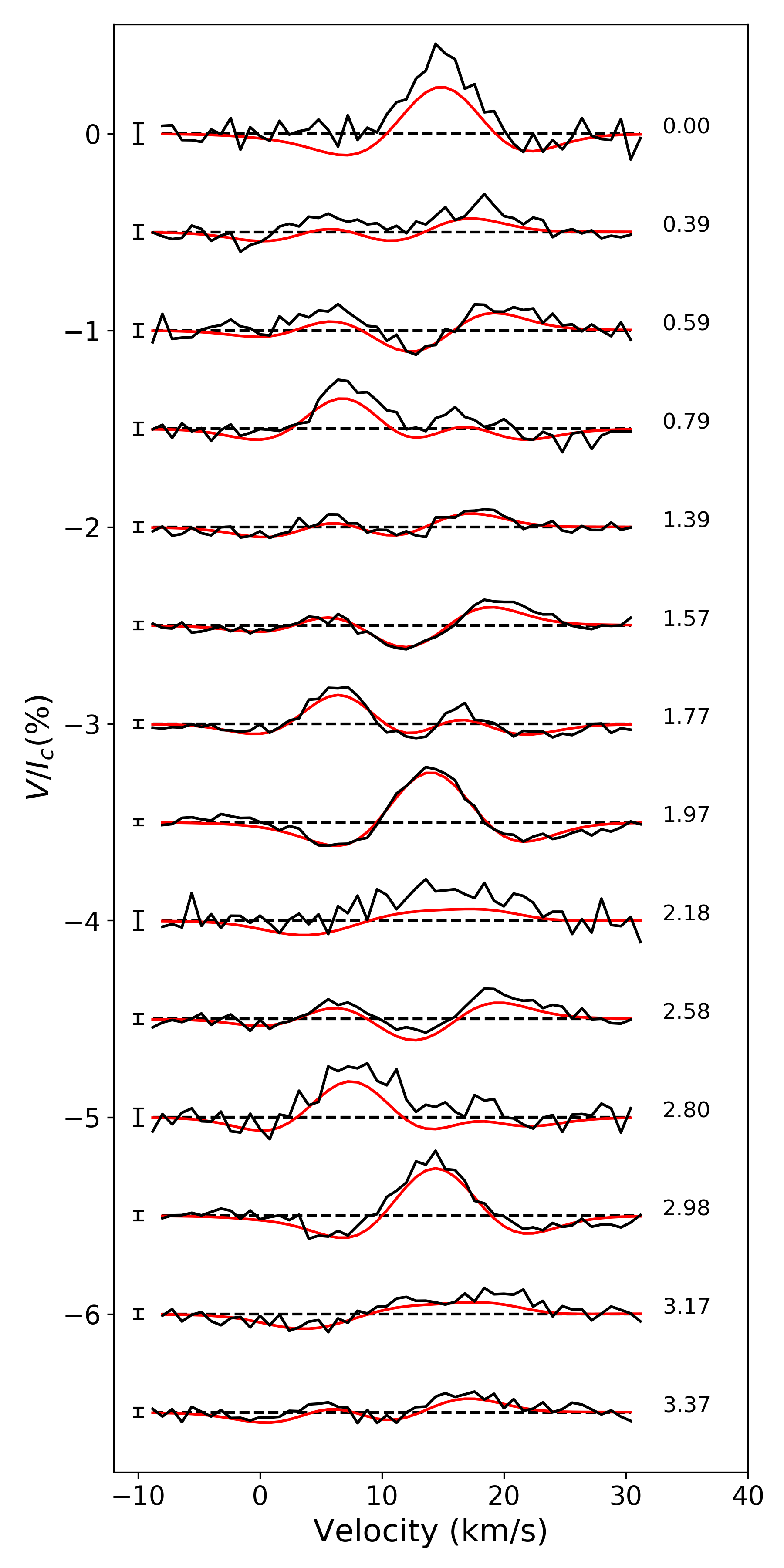}
\caption{Stokes V intensity, expressed as a percentage of Stokes I continuum of TWA 9A (shown in Figure \ref{fig:StokesI_TWA}). The black line is the observed data, and the red line the fit to the data with ZDI. Error bars on the left indicate the mean $1\sigma$ error for the observation, and the number on the right indicates the rotation phase of the observation based on 5.01 day rotation period. }
\label{fig:StokesV}
\end{figure} 

\subsection{Surface differential rotation}

The surface differential rotation of both stars was determined through the process of reconstructing the surface brightness map and, for TWA 9A, also using our large-scale magnetic field maps. Here we define differential rotation as: 
\begin{equation}
	\Omega(\theta) = \Omega_{\rm eq} - d\Omega\sin^2(\theta),
\end{equation}
where $\Omega(\theta)$ is the stellar rotation at latitude $\theta$, $\Omega_{\rm eq}$ is the equatorial angular velocity, and $d\Omega$ is the difference in rotation, or shear, between the equator and the pole. To determine the differential rotation from surface brightness we compute a grid of models, varying $\Omega$ and $d\Omega$, and fit for the minimum in the resulting $\chi^2_r$ surface with a 4th order paraboloid. The $3\sigma$ error is determined by the $3\sigma$ contour of the paraboloid. 

\begin{figure*}
\centering
\includegraphics[width=\textwidth]{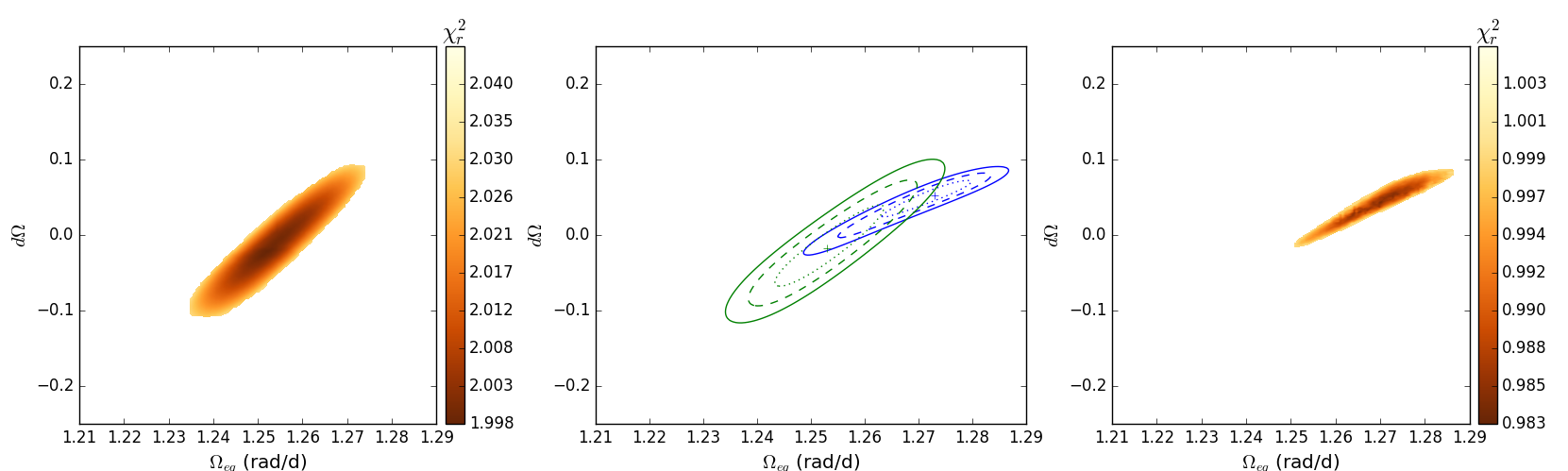}
\caption{Differential rotation measurements of TWA 9A from modelling of Stokes V (left) and Stokes I (right). Left: variations in $\chi^2_r$ as a function of equatorial rotation, $\Omega_{\rm eq}$, and shear, $d\Omega$ from circular polarisation profiles. Right: variations in $\chi^2_r$ as a function of $\Omega_{\rm eq}$ and $d\Omega$ from intensity profiles. Centre: paraboloid fits to the $\chi^2_r$ surface from Stokes I (blue) and Stokes V (green). The central plus indicates the best fit values, the dashed line the $1\sigma$ contour from the paraboloid fit, and the solid line the $3\sigma$ contour. }
\label{fig:diffrot_redchi3D_TWA}
\end{figure*}

Performing this analysis on the intensity profiles of V1095 Sco, we find no clear differential rotation solution. For TWA 9A, however, we find minima for both our brightness and magnetic field reconstructions. These are plotted in Figure \ref{fig:diffrot_redchi3D_TWA}, where each panel shows the $\chi^2_r$ surface as a function of $\Omega$ and $d\Omega$. The left-hand panel shows the $\chi^2_r$ surface from ZDI, the right-hand panel shows the $\chi^2_r$ surface from DI, and the center panel shows the paraboloid fits and minima  for both the DI (blue lines and plus) and ZDI (green lines and plus) maps. The DI map gives a solution of  $\Omega_{\rm eq}=1.273\pm0.007$ radians per day, and $d\Omega = 0.05 \pm 0.02$ radians per day, meaning the equator will lap the pole once every 125.7 days. The ZDI map, however, gives a shear solution of $\Omega_{\rm eq} = 1.254\pm0.008$ and $d\Omega = -0.013\pm0.013$, which could indicate anti-solar shear. Given the greater relative uncertainty on this value, we examine the overlapping region of our DI and ZDI $3\sigma$ elipses and take $d\Omega=0$ for our ZDI analysis.   

\subsection{Longitudinal magnetic field}
\label{sec:Bl}
\begin{figure}
  \centering
    \includegraphics[width=.5\textwidth]{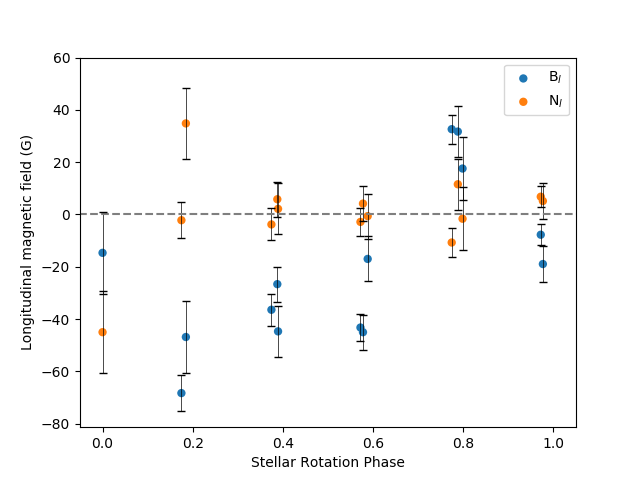}
  \caption{Longitudinal magnetic field for TWA 9A, calculated for each Stokes V profiles ($B_l$, blue circles), and in the Null profiles ($N_l$, orange circles), with $1\sigma$ uncertainties. }
  \label{fig:Bl_Nl_vs_phase_TWA9A}
\end{figure}

We also measure the line-of-sight, or longitudinal, magnetic field, $B_l$, for each Stokes V profile following \cite{Grunhut2013}, using the mean Land\'e factor and central wavelength of our LSD profiles. To determine the amount of spurious signal in these measurements, the same calculation is performed for each Null spectrum, denoted $N_l$ here.  For TWA 9A these values are plotted in Figure \ref{fig:Bl_Nl_vs_phase_TWA9A}. The $B_l$ values vary between $-68.3 \pm 6.9$ G and $32.6 \pm 5.5$ G. Only two $N_l$ values show spurious signal, although the associated $B_l$ values are still consistent with the rest of the sample.  The clustering of $B_l$ values at the same rotation phase indicate minimal changes to the large-scale field over the period of observations. A Generalised Lomb-Scargle period analysis of these values give a dominant period aligning with the stellar rotation period (see Section \ref{sec:periods} for more details). 

For V1095 Sco, we calculate the longitudinal field for each marginal or definite detection Stokes V profiles. These are not reliable, however, due to the high level of noise present, and these $B_l$ values are mostly the same order as the Null measurement. The observation of V1095 Sco with the strongest signal the least noise gives a longitudinal field of $25.9 \pm 16.0$ G. 

\section{Temporal variability of chromospheric emission}
\label{sec:halpha} 
Both stars in our sample are magnetically active, and therefore have strong chromospheric emission in, for example, the Balmer lines. We explore this by examining the \halpha ($\lambda 656.285$ nm) region of the stellar spectra to probe rotational modulation, and look for changes in activity over our period of observation. 

\begin{figure}
  \centering
    \includegraphics[width=.35\textwidth]{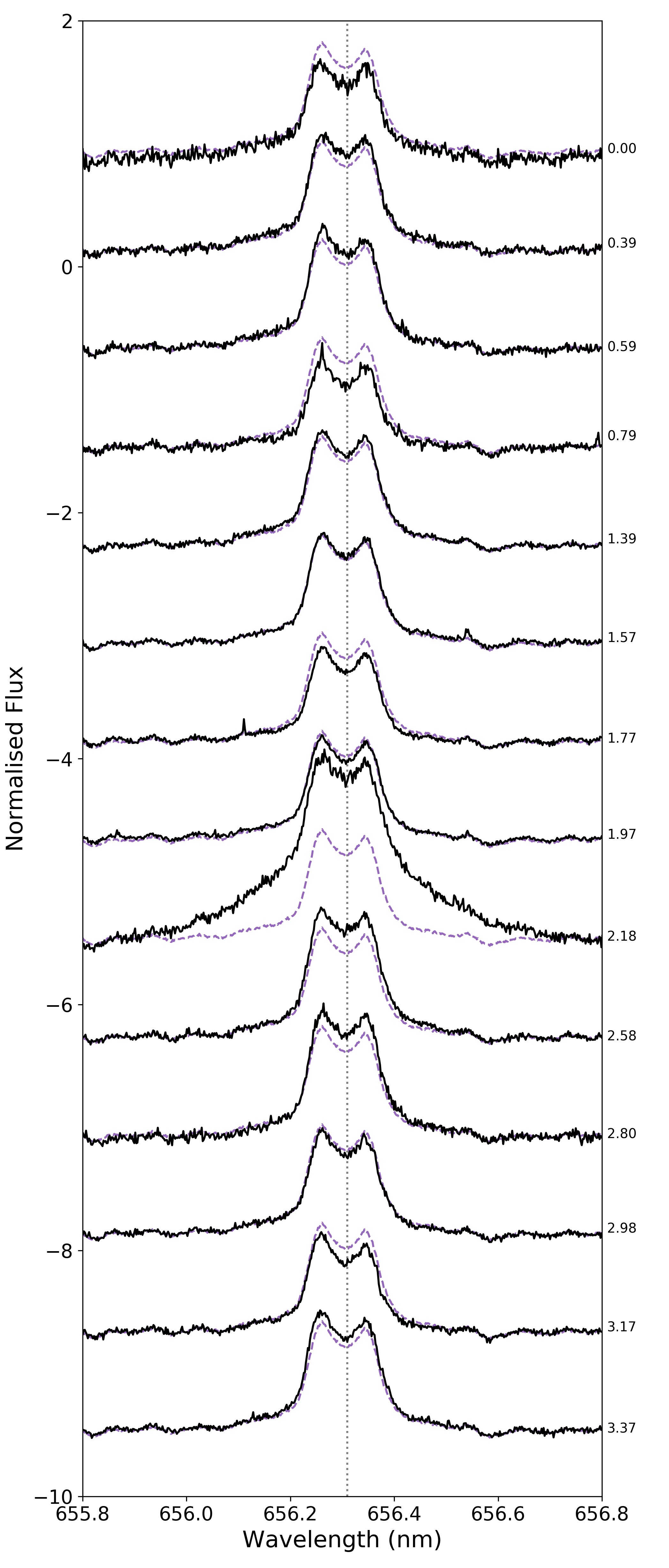}
  \caption{Stacked\halpha emission spectra for TWA 9A (black), plotted over the mean spectrum (purple, dashed), with the spectrum of the suspected flare excluded from the mean. Numbers on the right-hand side indicate the phase of observation. The grey dotted line indicated the centre of the\halpha line given the star's radial velocity. }
  \label{fig:TWA9A_Halpha_timeseries}
\end{figure}

For TWA 9A we find strong, double-peaked emission across all observations, with enhanced emission during the observations on 3 June 2014 (rotation phase 2.18). A plot of the stacked\halpha spectra for these observations is shown in Figure \ref{fig:TWA9A_Halpha_timeseries}. We see variation in the strength of the emission with stellar rotation, but minimal variation in the shape of the line, indicating the variation is likely due to areas of different activity level on the stellar surface. This emission is narrow and weak enough to fall below the accretion cut-off for a cTTS as defined by \cite{Cieza2013}, and therefore the emission is interpreted as chromospheric in nature. 

\begin{figure*}
  \centering
    \includegraphics[width=.9\textwidth]{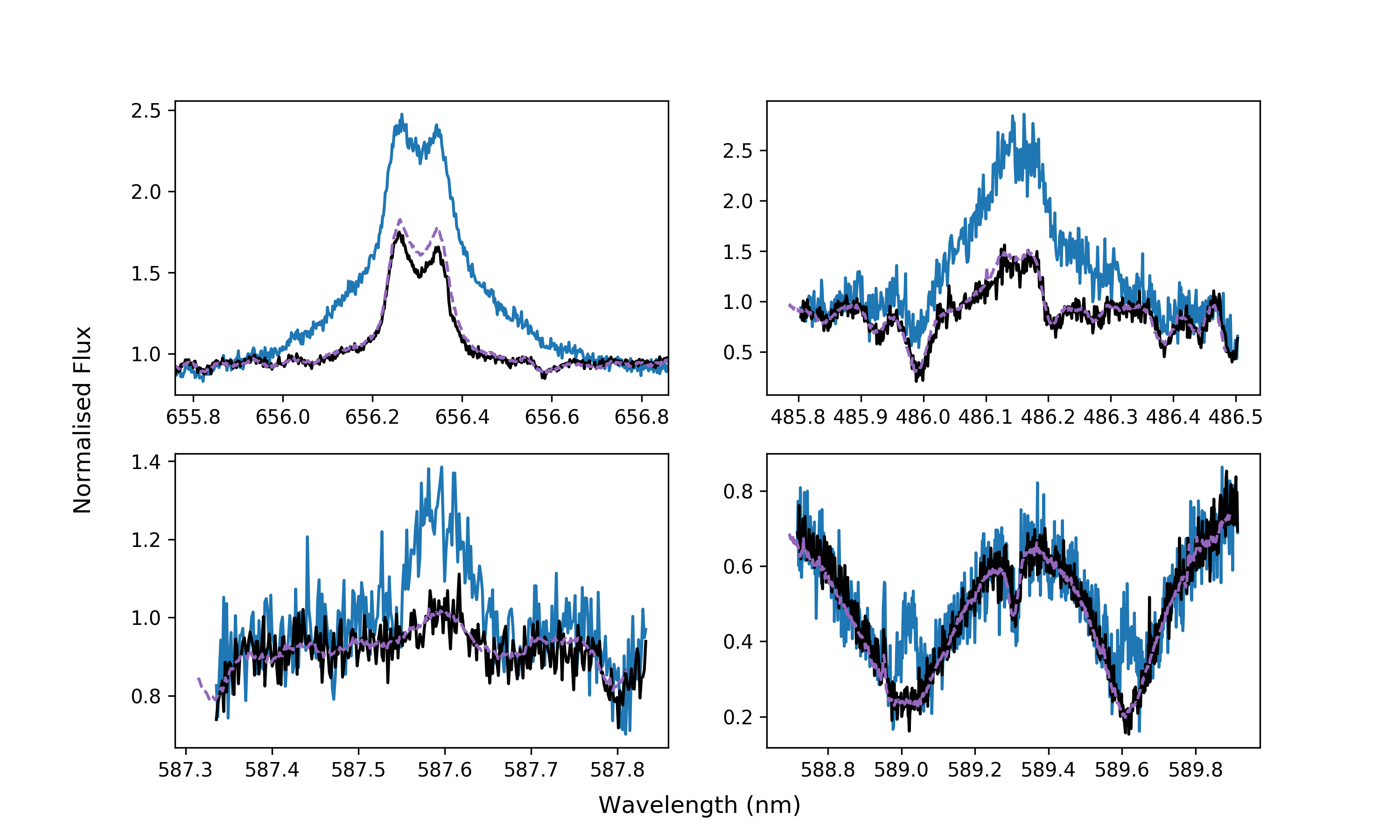}
  \caption{Emission in\halpha (top left), H$\beta$ (top right), He I (bottom left) and Na doublet (bottom right) of TWA 9A in the observation ascribed to a flare (blue), in an observation obtained one rotation later (black), and in the mean of the observed spectra (purple, dashed). }
  \label{fig:Flarespec_140602_HaHbHeINaI}
\end{figure*}

For the epoch with significantly enhanced\halpha emission (3 June 2014), we also examine the H$\beta$, He I and Na I doublet regions of the spectrum, and compare these regions to those taken one rotation later (phase 3.17), and to the mean spectrum. This is shown in Figure \ref{fig:Flarespec_140602_HaHbHeINaI}. Each of these regions exhibit different behaviour in the June 3 epoch, compared with that phase observed one rotation later and the mean spectrum. Both the HeI and Balmer lines show enhanced emission, and the Na doublet shows emission in the cores of the lines that is absent from the other observations. These changes in the spectral features, and the short timescales of the changes, suggest that this enhanced emisssion is most likely caused by a flare. 

Examining the Stokes I and V LSD profiles for that observation, we find no significant difference between that profile and the Stokes I and V profiles at same phase observed post-flare, so can conclude that this has had negligible influence on our tomographic modelling and radial velocity analysis. This is expected, as regions containing emission lines or strong absorption features, such as Na I, are excluded from our line list for the least squares deconvolution. 

\begin{figure}
  \centering
    \includegraphics[width=.35\textwidth]{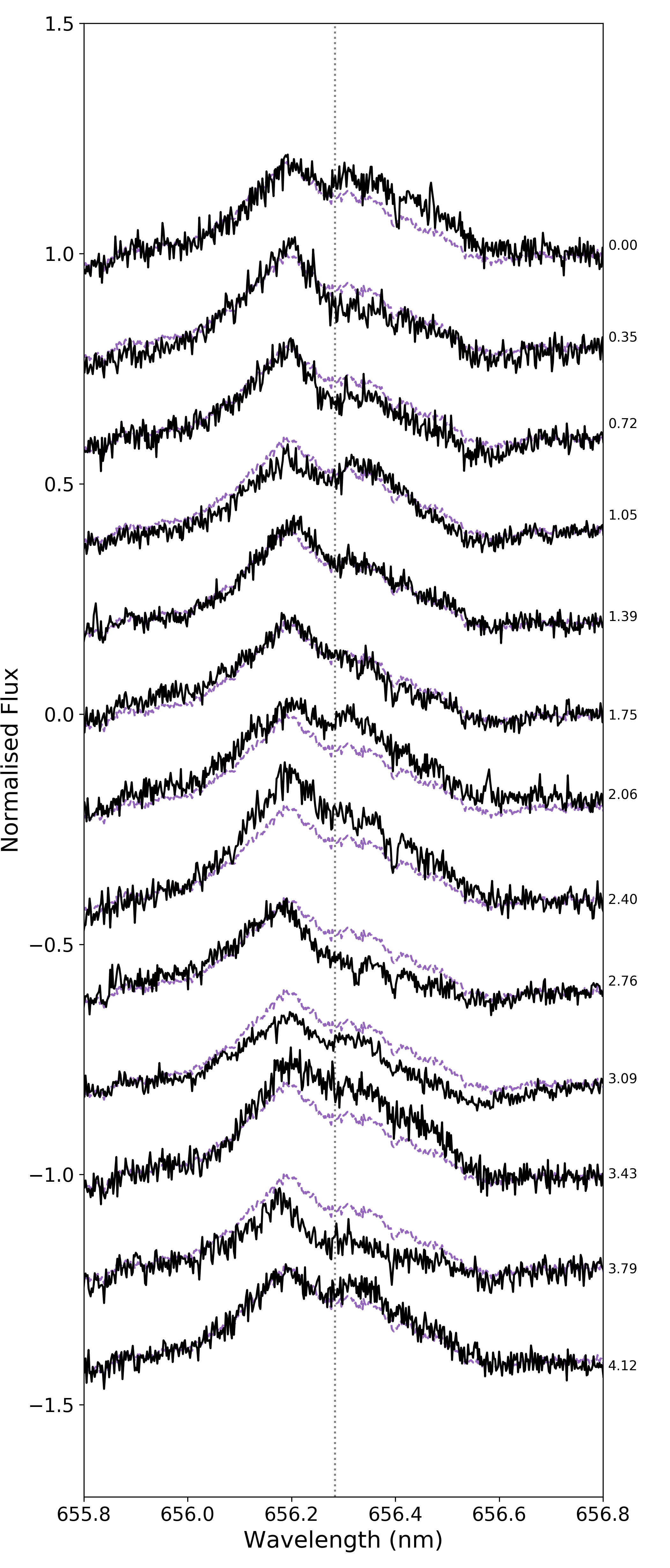}
  \caption{Stacked\halpha emission spectra for V1095 Sco (black), plotted over the mean spectrum (purple, dashed). Description as per Figure \ref{fig:TWA9A_Halpha_timeseries}. }
  \label{fig:V1095Sco_Halpha_timeseries}
\end{figure}

For V1095 Sco, a plot of the stacked\halpha spectra is shown in Figure \ref{fig:V1095Sco_Halpha_timeseries}. Overall, the emission is weaker than in TWA 9A.  We see enhanced emission on the blue-ward side of the profile (possibly indicating mass motion towards the observer) in a region around phase 0.4 and again in a region around phase 0.75, but more symmetric emission at phase 0.0 to 0.1.  

\begin{figure}
  \centering
    \includegraphics[width=.4\textwidth]{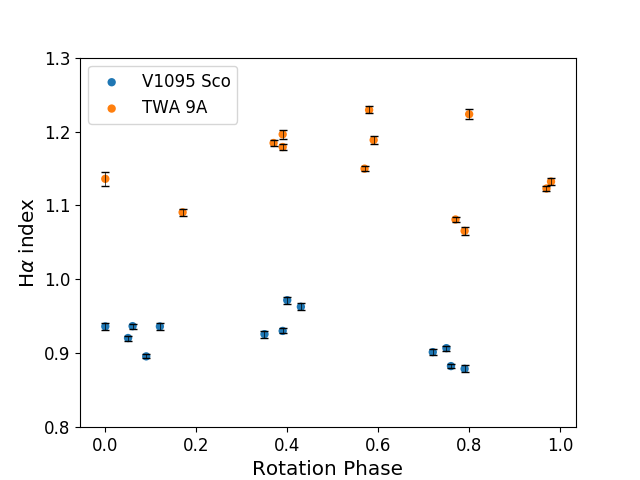}
  \caption{\halpha indices for V1095 Sco (blue) and TWA 9A (orange). Note that due to the large emission in the flare event of TWA 9A, that peak value of the\halpha index is underestimated due to saturation of the bandpass. }
  \label{fig:BothStars_Haindex_Dates_standard}
\end{figure}

We calculate an\halpha index for both stars as defined in Equation 9 of \cite{Marsden2014}, with a standard rectangular bandpass of 0.36 nm over the emission region, and two continuum band-passes of 0.22 nm centered on 655.885 nm and 656.730 nm. These measurements are shown for both stars in Figure \ref{fig:BothStars_Haindex_Dates_standard}, along with $3\sigma$ uncertainties. For V1095 Sco, whilst there does appear to be cyclical behaviour in the\halpha index, there also appears to be evolution over the period of observations, with an increase in the range of values in the later observations. For TWA 9A, the flare event is evident in the\halpha indices, with a significantly higher index for that observation compared to the rest, which seems to vary cyclically with stellar rotation (see Section \ref{sec:periods} for period analysis). It should be noted that due to the breadth of the emission in the flare spectrum, the standard bandpass is saturated, and the standard continuum regions chosen fall within the emission region for that observation, so that index value is likely underestimated. 


\section{Radial velocities and activity jitter}
\label{sec:RVs}
The process of reconstructing the brightness maps allows us to characterise and filter out the activity-induced jitter from the observed radial velocities \citep[see e.g.][]{Donati2015}. Our radial velocities are calculated as the first order moment of the Stokes I profile (i.e. $1-I$) as a function of velocity in the heliocentric rest frame. The radial velocity contribution of the stellar surface features is calculated as the radial velocity of our Doppler Imaging fits to the Stokes I LSD profiles. We then filter the activity jitter from the observed radial velocities by subtracting off the radial velocity values of these fits. The uncertainties for both the filtered and unfiltered radial velocities are calculated using the method of \cite{Butler1996}. 

\begin{figure}
\centering
\includegraphics[width=0.5\textwidth]{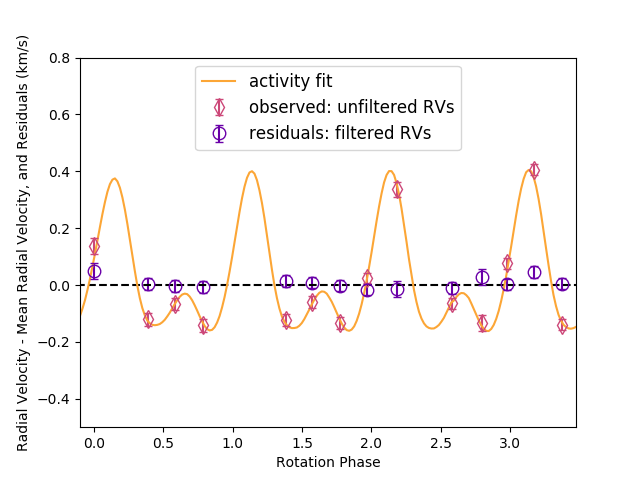}
\caption{Radial velocity analysis of TWA 9A. The pink diamonds are the measured radial velocities from the observed intensity profiles (RMS 187 \mps), the orange line is the modeled radial velocity contribution of the activity measurement from the surface brightness reconstruction, and the purple circles are the residuals after subtracting the observation radial velocities from the activity fit radial velocities (RMS 25 \mps).}
\label{fig:radialvelocityTWA9A}
\end{figure}

Analysis of the radial velocities of TWA 9A is shown in Figure \ref{fig:radialvelocityTWA9A}. The RMS of the unfiltered radial velocities is 187 \mps, and the activity filtered radial velocities have an RMS of 25 \mps, with the $1\sigma$ uncertainties averaging 20 \mps. Since the dispersion of our radial velocities is smaller than the $3\sigma$ uncertainty across all RV measurements, we are unable to detect any significant signal in the data (see Section \ref{sec:periods} for further period analysis). Additionally, this technique is insensitive to planets with periods equal to, or harmonics of, the stellar rotation period. With our current detection limits and RV measurements, we can rule out the presence of a planet with an $M_p \sin i$ greater than $\sim0.4$ Jupiter masses and closer than $\sim0.1$ au, assuming a circular orbit. 

\begin{figure}
\centering
\includegraphics[width=0.5\textwidth]{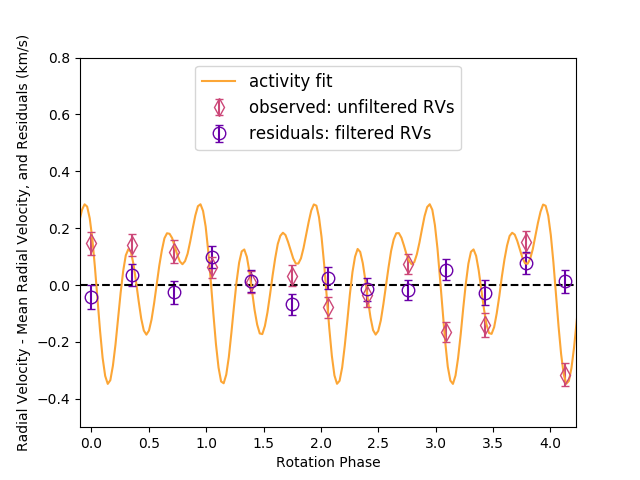}
\caption{Radial velocity analysis of V1095 Sco. The pink diamonds are the measured radial velocities from the observed intensity profiles (RMS 137 \mps), the orange line is the modeled radial velocity contribution of the activity measurement from the surface brightness reconstruction, and the purple circles are the residuals after subtracting the observation radial velocities from the activity fit radial velocities (RMS 47 \mps).}
\label{fig:radialvelocityV1095Sco}
\end{figure}

The same radial velocity analysis was performed for the radial velocities of V1095 Sco, shown in Figure \ref{fig:radialvelocityV1095Sco}. The RMS of the unfiltered radial velocities is 137 \mps, and $1\sigma$ uncertainties averaging 39 \mps. Applying the same filtering technique, we obtain an RMS of 47 \mps. Since the dispersion is smaller than the $3\sigma$ uncertainties, we are again unable to detect any significant signal in the data (see Section \ref{sec:periods} for further period analysis), but can rule out the presence of a planet with an $M_p \sin i$ greater than $\sim1.0$ Jupiter masses closer than $\sim0.1$ au, assuming a circular orbit. 

To explore the difference in using plage and spot DI versus the spot only version, we calculate the filtered radial velocities for TWA 9A and V1095 Sco using a spot-only model. For TWA 9A, the differences in the dispersion of the filtered radial velocities between the two models was insignificant $(<1\%)$. For V1095 Sco the dispersion of the plage and spot model filtered radial velocities was $\sim 9 \%$ smaller than the dispersion of the spot only model filtered radial velocities. 


\section{Periodograms of activity jitter and proxies}
\label{sec:periods}
\begin{figure}
  \centering
    \includegraphics[width=.5\textwidth]{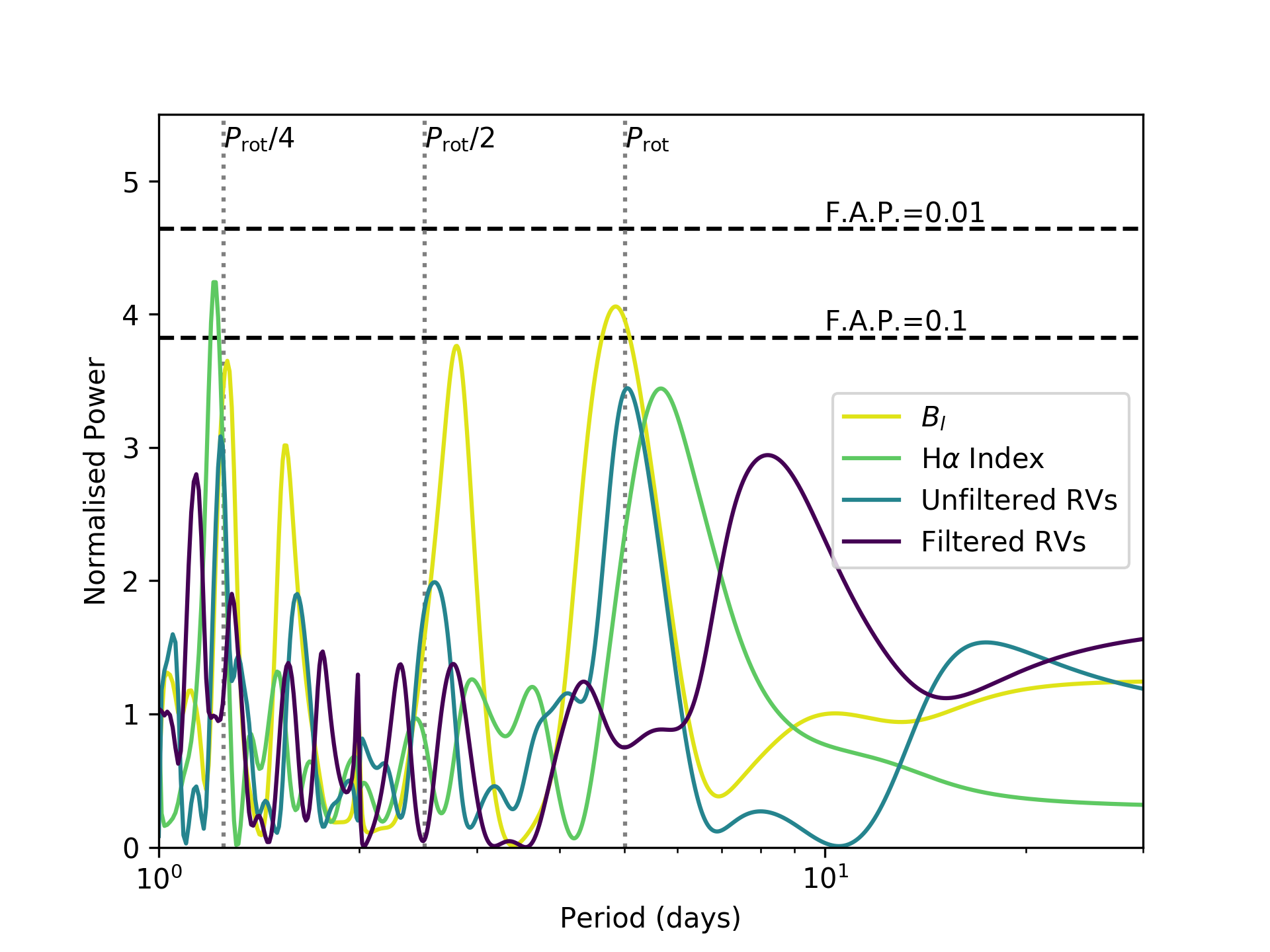}
  \caption{Generalised Lomb Scargle periodogram for TWA 9A of $B_l$ (yellow),\halpha (green), unfiltered radial velocities (blue), and filtered radial velocities (purple).  Dashed black lines indicate the F.A.P thresholds, and grey dotted lines indicate the 1, $1/2$ and $1/4$ multiples of the rotation period. }
  \label{fig:Periods_HaNoflare_Bl_RvInt_RvRes_TWA}
\end{figure}

There are a number of methods to search for rotational modulation in the multiple sets of diagnostics contained within our data sets to verify the rotation period determined from tomographic mapping. For both stars we examine the periodic nature of the\halpha emission (Section \ref{sec:halpha}), radial velocities, both unfiltered and filtered (Section \ref{sec:RVs}), and for TWA 9A the $B_l$ values (Section \ref{sec:Bl}). For TWA 9A we exclude the possible flare event from the\halpha index data set for our period analysis. We perform a period analysis with a Generalised Lomb Scargle periodogram \citep{Zechmeister2009}, using the PyAstronomy\footnote{\url{https://github.com/sczesla/PyAstronomy}} package. The periodograms for TWA 9A are shown in Figure \ref{fig:Periods_HaNoflare_Bl_RvInt_RvRes_TWA}, including false alarm probability (F.A.P) thresholds of 0.1 and 0.01. The F.A.P values are calculates as per Equation 24 of \cite{Zechmeister2009}. In each of the\halpha, $B_l$ and unfiltered radial velocities, there is a peak in the power around our DI-determined stellar rotation period. The signal is clearest in the longitudinal field measurements compared to the\halpha indices whose peak is shifted to slightly longer periods, likely due to this diagnostic probing activity at higher altitudes. The stellar rotation period is absent from the filtered radial velocities, indicating that the jitter filtering process is effective at removing the first order stellar rotation signal.

\begin{figure}
  \centering
    \includegraphics[width=.5\textwidth]{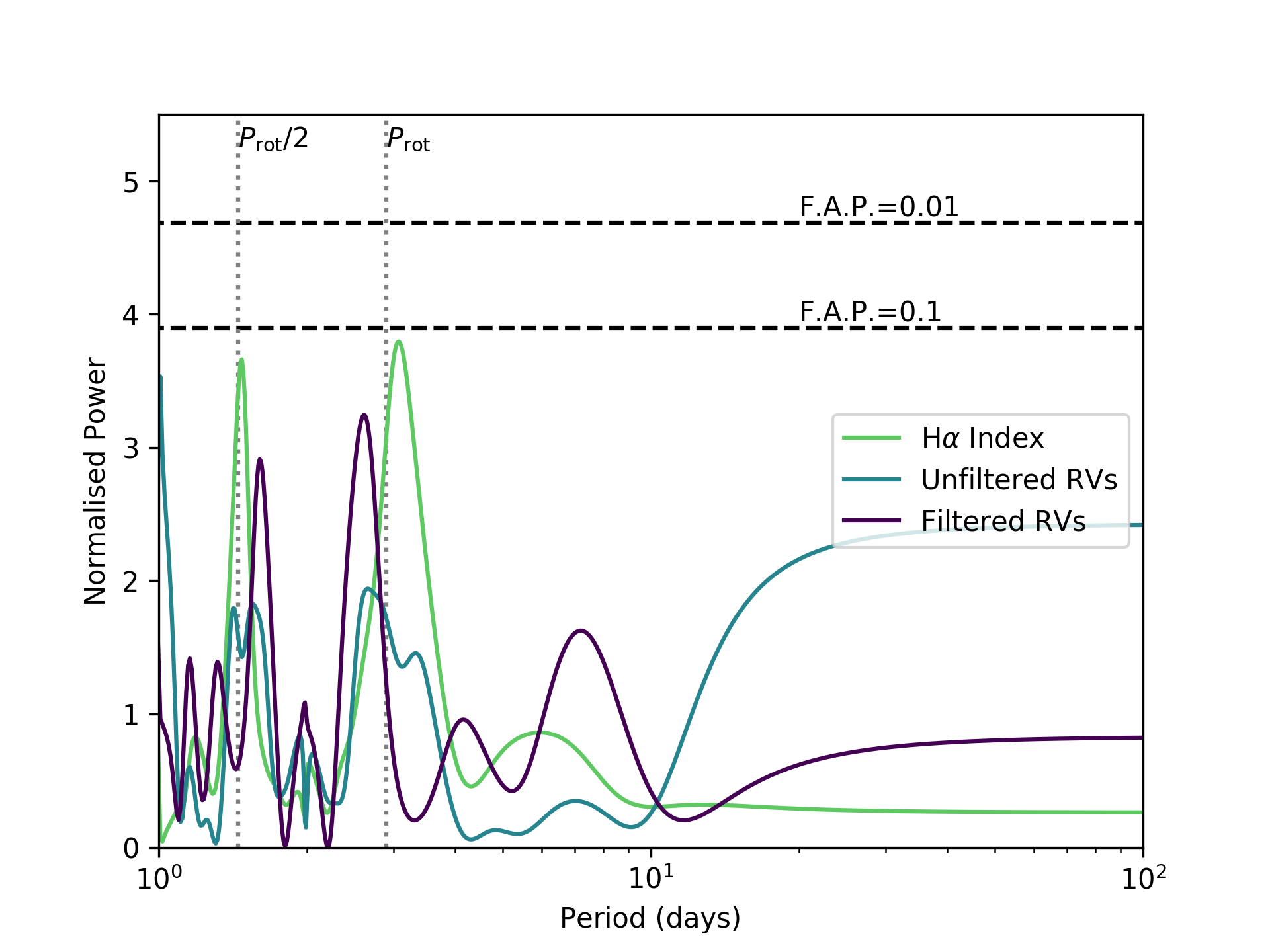}
  \caption{Generalised Lomb Scargle periodogram for V1095 Sco of\halpha (green), unfiltered radial velocities (blue), and filtered radial velocities (purple).  Dashed black lines indicate the F.A.P thresholds, and grey dotted lines indicate the 1 and $1/2$ multiples of the rotation period.}
  \label{fig:Periods_Ha_RVs_RXJ}
\end{figure}

The periodogram for V1095 Sco is shown in Figure \ref{fig:Periods_Ha_RVs_RXJ}. The periodicity in the unfiltered and filtered radial velocities is less clear than in the case of TWA 9A, likely resulting from the larger uncertainties in the radial velocity measurements and the poorer sampling, with the periods with the highest peaks being below our sample rate of $\sim 1$ day$^{-1}$. As such, this diagnostic cannot be used to determine the effectiveness of our radial velocity filtering for this star.  Nevertheless, the\halpha indices do show a peak of highest power around the DI determined rotation period, supporting our DI rotation period measurement for V1095 Sco.


\section{Summary and Discussion}
\label{sec:discussion}

This paper presents the results of spectropolarimetric observations of two wTTSs, TWA 9A and V1095 Sco, as part of the MaTYSSE Large Programme. These stars are studied as a pair as they represent two stages, pre and post radiative core formation, in 1\Msun wTTS evolution. Observations for TWA 9A spanned 18 days, and for V1095 Sco spanned 13 days, and were obtained using the HARPS spectrograph on the ESO 3.6m Telescope at La Silla Observatory, Chile. The techniques of Doppler Imaging (DI) and Zeeman Doppler Imaging (ZDI) were applied to both data sets to reconstruct the large-scale surface brightness and magnetic field features. From the surface activity information from the DI, we were able to filter the activity jitter from the radial velocities of both stars. This enabled radial velocity measurements that rule out the presence of close-in giant planets around these stars. We also examined the strongest of the Balmer emission lines,\halpha, as a source of information on the variability of chromospheric activity. We performed a period analysis on the\halpha activity index in combination with radial velocities (and longitudinal field in the case of TWA 9A) to verify the DI\RCadd{-}determined rotation period, and demonstrate the effectiveness of the radial velocity filtering process in removing the stellar activity signal.

\subsection{Dynamos of young, sun-like stars}

TWA 9A and V1095 Sco have very different brightness maps, even when taking their different \vsini values and therefore different spatial resolution of the maps into account. This may imply a difference in the underlying flux emergence patterns of these stars. The brightness map of TWA 9A shows evidence of both cool spots and bright plage, and resembles the topology of V819 Tau \citep{Donati2015} and Par 1379 \citep{Hill2017}, which are both younger than TWA 9A, but have similar rotation periods and masses. V1095 Sco's map is dominated by a large cool polar spot, with weak evidence of plage on its surface. This is in contrast to all other wTTSs mapped in the MaTYSSE sample so far, which all exhibit plage signatures on their surfaces. However, the large polar spot, rapid rotation, poorer SNR, and poorer phase coverage of V1095 Sco make it unclear if this star is truly an outlier in the MaTYSSE sample. Of particular contrast to V1095 Sco is V830 Tau, which exhibits a more complex brightness topology, despite having a similar age, mass, \vsini, rotation period  and internal structure. This difference can potentially be explained by shorter time-scale magnetic field evolution in these types of stars, and were they to be observed at a different epoch, might display similar characteristics. V1095 Sco's surface activity does resemble that of AB Doradus, a 50 Myr old, rapidly rotating post-T Tauri star, which displays a large polar spot in addition to some lower attitude spot features \citep{DonatiCC1997,Donati2003,Hussain2007}. 

The reconstructed magnetic field topology of TWA 9A is predominantly poloidal ($69\%$ poloidal) and  non-axisymmetric ($57\%$ non-axisymmetric), with the dipolar component tilted at 36 degrees from the rotation axis. The overall field has a mean strength of 113 G and a maximum strength of 296 G. This resembles the topology and complexity of V819 Tau and Par 1379, but has the weakest intensity of dipole field of any mapped in the MaTYSSE sample so far (V1095 Sco likely has a weaker field, but its magnetic field has not been mapped). Comparing to the MAPP sample of cTTS, TWA 9A is of a similar age and mass to V4046 Sgr A \citep{Donati2011}, and displays a similar level of axisymmetry, but again has a weaker field strength, and a more poloidal field.  Comparing to a wider sample of young ($<250$ Myrs) stars from the TOUPIES (TOwards Understanding the sPIn Evolution of Stars) project presented by \cite{Folsom2016}, TWA 9A has magnetic field characteristics within the range seen in that sample, despite it being $\sim 10$ Myrs younger than the youngest stars in that sample.  This is consistent with the notion that rotation speed and internal structure play a dominant role in determining the magnetic, and hence surface, activity for young stars \citep{Folsom2016, Gregory2016}. 

The longitudinal magnetic field was calculated for each observation of TWA 9A, giving values ranging from $-68.3 \pm 6.9$ G to $32.6 \pm 5.5$ G, and varying over a period coinciding with the stellar rotation period. The longitudinal field was measured for the strongest and most reliable observation of V1095 Sco, giving a value of $25.9 \pm 16.0$ G. There are only 2 wTTSs in the literature with published longitudinal field measurements, namely TAP 26 \citep{Yu2017} and V410 Tau \citep{Skelly2010}. Both have far higher maximum $B_l$ values than either of the stars presented here, which also correlates with those stars having greater large-scale magnetic field strengths. 

Differential rotation was investigated for both stars, but a solution was only found for TWA 9A. The intensity profiles of TWA 9A indicate strong differential rotation with an equatorial rotation of $\Omega_{\rm eq}=1.273\pm0.007$ radians per day, and shear of $d\Omega = 0.05 \pm 0.02$ radians per day. This is very similar to the solar shear value of $\sim 0.055$ radians per day, and is the largest Stokes I differential rotation measurement of the published MaTYSSE sample to date (though only marginally higher than that of LkCa 4 \citep{Donati2014}). 

In both stars we observe emission in the\halpha region of their spectra as a diagnostic of their chromospheric activity. The emission is weaker for V1095 Sco than for TWA 9A, which is consistent with the idea of a weaker magnetic field in the former star. The shapes of the profile are different in each case, with V1095 Sco showing variation in the shape of the line profile, and excess emission on the blue-ward side of the\halpha region, indicative of mass motion toward the observer. In contrast, TWA 9A shows symmetric, double peaked emission that varies cyclically in intensity, with the exception of a possible flare event on 3 June that exhibits far higher and broader emission. 

In overall terms, the observed behaviour of both stars is consistent with intense magnetic activity. Comparing to the\halpha emission published for other stars in the MaTYSSE sample, namely Par 1379, Par 2244 and V410 Tau, the emission of V1095 Sco is the weakest, and V1095 Sco is the only one to exhibit excess emission on the blue-ward side of the profile. TWA 9A has emission intensity on the same order as the other MaTYSSE stars, but does not show the same level of rotational and temporal variability of intensity, or profile shape. Comparing to the\halpha emission of TWA 17 published by \cite{Skelly2009}, a star of the similar age, mass and spectral type, we see on average weaker emission for TWA 9A, which is likely explained by the slower rotation of TWA 9A leading to comparatively lower activity. The broad `pedestal' emission (emission in the line far outside the rotational broadening limits) seen in the mean\halpha of TWA 17, which is linked to the presence of microflaring \citep[e.g.][]{Fernandez2004}, is absent from the\halpha emission of TWA 9A, but is seen in the enhanced emission on June 3, further supporting the idea that this enhanced emission has resulted from a flaring event.

\subsection{Exoplanets around young stars}

In addition to the study of wTTS dynamos, the other aim of the MaTYSSE Programme is the search for young, close-in, giant exoplanets. Here, we analyse the radial velocity curves of both stars to search for a planet-induced radial velocity signal. Since both stars exhibit strong surface activity, and hence activity jitter in their radial velocity curves, we use our DI analysis to characterise and remove this jitter. By doing this we reduce the radial velocity RMS in V1095 Sco from 137 \mps to 48 \mps, and in TWA 9A from 187 \mps to 25 \mps. No signal is found in the filtered radial velocity curve, allowing us to rule out the presence of a $\sim1.0$ Jupiter mass planet closer than $\sim0.1$ au in the case of V1095 Sco, and a planet larger than $\sim0.4$ Jupiter masses and closer than $\sim0.1$ au for TWA 9A. This difference in detection limits for the two stars is likely due to the differences in rotation rates and quality of the respective data sets. V1095 Sco has a\vsini nearly three times that of TWA 9A, so has greater Doppler broadening of its spectra, which decreases the precision of the radial velocity measurements. In addition to that, V1095 Sco's data have a lower signal-to-noise ratio, which not only impacts the precision of the radial velocity measurements themsleves, but also affects the quality of the DI fits. This, in turn, results in a poorer filtering of the activity jitter, leading to less of a reduction in radial velocity RMS in V1095 Sco compared to TWA 9A. 

Of the 8 MaTYSSE stars studied so far, there have been 2 exoplanet detections, one each for V830 Tau \citep{Donati2015,Donati2016,Donati2017} and TAP 26 \citep{Yu2017}. This puts the occurrence rate of close-in hot Jupiters in the MaTYSSE sample at 1 in 4, which appears remarkably high in comparison to the $\sim 1\%$ occurrence rate in MS stars from transit and radial velocity surveys \citep{Guo2017}. In addition to the planets found by MaTYSSE, \cite{David2016} found a close-in Neptune-sized planet around a 5-10 Myr old star, detected through the transit method. Given these independent detections of young close-in planets, further work is needed to understand their formation and eventual fate as their host stars evolve onto the main sequence.


%
%
%

\section*{Acknowledgements}
This work has made use of the VALD database, operated at Uppsala University, the Institute of Astronomy RAS in Moscow, and the University of Vienna. This work has made use of data from the European Space Agency (ESA) mission {\it Gaia} (\url{http://www.cosmos.esa.int/gaia}), processed by the {\it Gaia} Data Processing and Analysis Consortium (DPAC, \url{http://www.cosmos.esa.int/web/gaia/dpac/consortium}). Funding for the DPAC has been provided by national institutions, in particular the institutions participating in the {\it Gaia} Multilateral Agreement. We also warmly thank the IDEX initiative at Universit\'e F\'erale Toulouse Midi-Pyr´ ees (UFTMiP) for funding the STEPS collaboration programme between IRAP/OMP and ESO and for allocating a ‘Chaire d'Attractivit\'e’ to GAJH, allowing her to regularly visit Toulouse to work on MaTYSSE data. This research is supported by USQ's Strategic Research Initiative programme.




\bibliographystyle{mnras}
\bibliography{twa9a_bib} 



%
%



\bsp	
\label{lastpage}
\end{document}